\documentclass[prd,twocolumn,superscriptaddress,showpacs,nofootinbib,amsmath,amssymb]{revtex4}

\usepackage{bm}
\usepackage{amsfonts}
\usepackage{latexsym}
\usepackage[latin1]{inputenc}
\usepackage{graphicx}
\usepackage{amsmath}
\usepackage{rotating}
\usepackage{epsfig}

\newcommand{\mb}[1]{\mbox{\boldmath $#1$}}

\def \nn  {\nonumber}

\def\jnl@style{}
\def\aaref@jnl#1{{\jnl@style#1}}

\def\aaref@jnl#1{{\jnl@style#1}}

\def\aj{\aaref@jnl{AJ}}                    
\def\apj{\aaref@jnl{Astrophys.~J.}}        
\def\apjl{\aaref@jnl{Astrophys.~J.~Lett.}} 
\def\apjs{\aaref@jnl{ApJS}}                
\def\apss{\aaref@jnl{Ap\&SS}}              
\def\aap{\aaref@jnl{Astron.~Astrophys.}}   
\def\aapr{\aaref@jnl{A\&A~Rev.}}           
\def\aaps{\aaref@jnl{A\&AS}}               
\def\mnras{\aaref@jnl{Mon. Not. R. Astron. Soc.}}             
\def\prd{\aaref@jnl{Phys.~Rev.~D}}        
\def\prl{\aaref@jnl{Phys.~Rev.~Lett.}}    
\def\qjras{\aaref@jnl{QJRAS}}             
\def\skytel{\aaref@jnl{S\&T}}             
\def\ssr{\aaref@jnl{Space~Sci.~Rev.}}     
\def\zap{\aaref@jnl{ZAp}}                 
\def\nat{\aaref@jnl{Nature}}              
\def\aplett{\aaref@jnl{Astrophys.~Lett.}} 
\def\apspr{\aaref@jnl{Astrophys.~Space~Phys.~Res.}} 
\def\physrep{\aaref@jnl{Phys.~Rep.}}      
\def\physscr{\aaref@jnl{Phys.~Scr}}       

\begin{document}

\title[Nonlinear] {Gravitational Waves from Nonlinear Couplings of
  Radial and Polar Nonradial Modes in Relativistic Stars}

\author{Andrea Passamonti}
\affiliation{Department of Physics, Aristotle University of
  Thessaloniki, 54124 Thessaloniki, Greece}
\author{Nikolaos Stergioulas}
\affiliation{Department of Physics, Aristotle University of
  Thessaloniki, 54124 Thessaloniki, Greece}
\author{Alessandro Nagar}
\affiliation{Dipartimento di Fisica, Politecnico di Torino, Corso Duca 
degli Abruzzi 24, 10129 Torino, Italy \\
and INFN, sez. di Torino, Via P. Giuria 1, Torino, Italy}

\date{\today}

\begin{abstract}
  The post-bounce oscillations of newly-born relativistic stars are
  expected to lead to gravitational-wave emission through the
  excitation of nonradial oscillation modes. At the same time, the
  star is oscillating in its radial modes, with a central density
  variation that can reach several percent. Nonlinear couplings
  between radial oscillations and polar nonradial modes lead to the
  appearance of combination frequencies (sums and differences of the
  linear mode frequencies). We study such combination frequencies
  using a gauge-invariant perturbative formalism, which includes
  bilinear coupling terms between different oscillation modes.  For
  typical values of the energy stored in each mode we find that
  gravitational waves emitted at combination frequencies could become
  detectable in galactic core-collapse supernovae with advanced
  interferometric or wide-band resonant detectors.

\end{abstract}


\maketitle

%
\section{Introduction}
\label{Sec:intro}
%

Oscillations of neutron stars can arise in strongly nonlinear
phases of stellar evolution, such as during the post-bounce phase in
core collapse of massive stars or during the merger of compact binary
systems.  Linear perturbation theory is appropriate for describing the
dominant properties of stellar pulsations, but neglects several
interesting nonlinear effects that can modify or enrich the linear
description.  For instance, a nonlinear approach is required for
understanding how energy transfer between various classes of modes
could saturate the ${\textrm r}$- and ${\textrm f}$-mode instabilities
in rotating stellar models (see~\cite{KS:2005pr} and references therein), or
limit the persistence of the bar-mode instability, through nonlinear
coupling between $m=1$ and $m=2$ modes~\cite{2006astro.ph..9473B, 2003ApJ...595..352S, 2005ApJ...625L.119O}.
Nonlinear mode couplings also become important in the development of a
${\textrm g}$-mode instability, that was recently observed in 
core-collapse simulations~\cite{Ott:2005}. Another nonlinear effect relevant
for pulsations in rapidly rotating stars is the mass-shedding-induced
damping of pulsations~\cite{Stergioulas:2003ep,Dimmelmeier:2005zk}.

The main nonlinear effect appearing even for mildly nonlinear
pulsations is the presence of combination tones, which are nonlinear
harmonics whose frequency appears as a linear combination of linear
eigenfrequencies (see ~\cite{Dimmelmeier:2005zk, 2005MNRAS.356.1371Z, 
  Passamonti:2005cz}). The number of linear terms in the combination
tones is equal to the nonlinear order of the mode coupling. In perturbation
theory, these nonlinear features can be addressed starting at bilinear
order, for the coupling of two linear modes.
 
The presence of nonlinear harmonics can lead to exchange of energy
between modes or even to two-mode or three-mode resonances or
parametric instabilities~\cite{Passamonti:2005cz}.  In rotating
proto-neutron stars, rotational effects can increase the number of
possible resonances between axisymmetric modes~\cite{Dimmelmeier:2005zk}. 
The parameter space for possible resonances
or instabilities is even larger for non-axisymmetric oscillations. In
this case, each $l$-mode is split in its $m$ components proportionally
to the rotation rate of the star. Therefore, at large rotation rates, 
several modes could satisfy resonance conditions~\cite{KDK:per}.

Gravitational waves emitted by neutron star pulsations are interesting
sources for current detectors, but the high-frequency band will only
start becoming accessible with the construction of advanced detectors,
such as Advanced LIGO~\cite{ligo}, VIRGO+ and Advanced
VIRGO~\cite{Ad_VIRGO_site,WP-Adv-VIRGO}. There are also plans for
detectors with significantly increased sensitivity at several kHz,
such as the GEO-HF project~\cite{Willke:2006uw}, and the proposal for
the Dual wide-band bar detector~\cite{2003PhRvD..68j2004B}. Nonlinear
effects in neutron star pulsations could be probed with such
instruments.

Interest in the nonlinear dynamics of stellar pulsations exists also
for variable stars, such as Cepheids, RR Lyr$\ae$ and $\delta$ Scuti
stars. The modulation of the pulsation amplitude (Blazkho effect) or
other irregular oscillations, which have been observed in the velocity
and light curve of these stars, could be explained by nonlinear
interaction between various modes. These studies have been carried out
mainly in Newtonian perturbation theory by using amplitude equations
(see e.g.~\cite{2003A&A...401..661K, 2004AcA....54..363D,
  2005AcA....55....1N} and references therein).

General Relativity provides a more appropriate framework for studying
compact objects, where relativistic effects, such as the dragging of
the inertial frames or the dynamical role of the spacetime, can
influence the spectral properties of hydrodynamical modes or introduce
new features, such as $w$-modes~\cite{kokkotas-1999-2}. Modern
high-resolution methods~\cite{{2003LRR.....6....4F}} (in particular,
the 3rd-order PPM scheme) implemented in numerical relativity codes,
have allowed in recent years the study of nonlinear pulsations in
nonrotating and rotating stars 
(see~\cite{FSK,FDGS,Font:2001ew,Stergioulas:2003ep,Dimmelmeier:2005zk}).
The above studies included 2-D and 3-D simulations of axisymmetric modes
in either the relativistic Cowling approximation, the 
spatially conformal flatness  (CFC) approximation or full general
relativity. Nonlinear relativistic radial pulsations
have also been addressed with a different method
in~\cite{Sperhake:2001si, Sperhake:2001xi}, where the nonlinear
dynamics was studied as a deviation from an equilibrium background.

Complementary to fully nonlinear approaches, nonlinear perturbative
techniques can be very useful in weakly and mildly nonlinear regimes
for understanding how a physical process is affected by nonlinear
effects.  In addition, the dimensional reduction of the problem by
separation of variables (when possible) yields a much less
computationally expensive approach, suitable for large parameter
studies.  In order to correctly address the gauge freedom of general
relativity, a relativistic perturbation theory has been introduced
in~\cite{Bruni:1996im}, and then extended to physical systems where
two or more variables can be assumed as perturbative
parameters~\cite{Bruni:2002sm, Sopuerta:2003rg, 2003PThPh.110..723N}. 
For spherically symmetric and time dependent spacetimes, a gauge
invariant formalism for studying spherical and nonspherical
perturbations was introduced by Gerlach and
Sengupta~\cite{Gerlach:1979rw, Gerlach:1980tx}. This formalism was
further developed in~\cite{Gundlach:1999bt, Martin-Garcia:2000ze} and
then extended (with partial gauge-invariance at second and higher
perturbative orders) in~\cite{2006PhRvD..74d4039B} (from now on GSGM
formalism). Efforts are now under way for formulating a fully
gauge-invariant theory at second or higher
order~\cite{2006PhRvD..74d4039B}.

Second-order perturbative studies have been carried out in
cosmology~\cite{Bruni:2001pc}, or for investigating the
Schwarzschild~\cite{Gleiser:1995gx, 1999bhgr.conf..351P, Garat:2000gp}
and Kerr~\cite{Campanelli:1998jv} spacetimes.
In the two-parameter relativistic perturbative framework, the coupling
between the radial and nonradial oscillations of perfect-fluid,
spherically symmetric neutron stars has been recently investigated
in~\cite{Passamonti:2004je, Passamonti:2005cz, passamonti-2006}. In
particular, the coupling between the radial and axial
oscillations exhibits an interesting resonance when a radial mode
frequency approaches the frequency of an axial
$w$-mode~\cite{Passamonti:2005cz, passamonti-2006}.  

In this paper, we address the coupling between the radial and polar
nonradial modes by numerically solving the perturbation equations
introduced in~\cite{Passamonti:2004je} for polytropic, nonrotating
relativistic stars. In the polar sector, new interesting results are
expected as the spectrum of polar modes is generally richer than the
axial case.  In particular, we set up initial conditions (in following
recent numerical simulations of core
collapse~\cite{Dimmelmeier:2002bk, Dimmelmeier:2002bm}) for studying
pulsating protoneutron stars.  In the post-bounce phase a newly-born
neutron star is expected to mainly oscillate in its radial ($l=0$) and
nonradial quadrupole ($l=2$) modes~\cite{Dimmelmeier:2002bk,
Dimmelmeier:2002bm} with a variation in its central density of several
percent.  After investigating several initial pulsating
configurations, we have found that for a $5\%$ variation in the
central energy density and about $10^{-7} M_{\odot}$ energy stored in
nonradial oscillations some bilinear combination tones are within the
sensitivity window of advanced and newly proposed gravitational wave
detectors. A possible detection of these combination tones, in
addition to the linear modes, would provide new important information
for solving the high-density equation of state puzzle through
gravitational-wave asteroseismology.
Note that the perturbative equations as developed in~\cite{Passamonti:2004je,
Passamonti:2005cz, passamonti-2006} do not contain back-reaction
terms. This issues is the subject of future work.

The plan of this paper is as follows. Sections~\ref{Sec:equil}
and~\ref{Sec:framework} are dedicated to the description of the
background stellar configuration and the perturbative framework,
respectively. The set of perturbed initial data, boundary conditions
and main properties of the numerical code are addressed in
Sec.~\ref{Sec:numcode}.  In Sec.~\ref{Sec:results}, we present
the temporal and spectral properties of the fluid and Zerilli variables,
which arise from the coupling between radial and polar nonradial
oscillations. The detectability of the gravitational wave signal is
discussed in Sec.~\ref{Sec:Strain}.
In Appendix~\ref{sec:non-radial-eqs} we present the equations that 
complete the system of nonradial perturbation equations given 
in~\cite{Passamonti:2004je}, while in Appendix~\ref{sec:Kin_Energy} 
we derive the pulsational kinetic energy of nonradial oscillations 
in terms of GSGM formalism.

In this paper the geometrical units are adopted, where $G = c =
1$.

%
\section{Background Equilibrium \label{equilibrium}}
\label{Sec:equil}
%

The background spacetime is the equilibrium configuration of a
perfect-fluid spherically symmetric nonrotating star, which is
described by the following spacetime metric:
\begin{equation}
ds^2= -e^{2\Phi} dt^2+e^{2\Lambda} dr^2+
r^2(d\theta^2+\sin^2\theta d\phi^2)\,,
\end{equation}
where $\Phi=\Phi(r)$ and $\Lambda=\Lambda(r)$ are obtained by solving
the Tolman-Oppenheimer-Volkov (TOV) equations. This system of 
equations is
closed by an equation of state (EoS) that characterizes the fluid
properties. In this paper we consider a relativistic barotropic EoS:
\begin{eqnarray}
p & = & K \rho ^{\Gamma} \, ,\\
\varepsilon & = & \rho + \frac{p}{\Gamma -1} \, ,
\end{eqnarray}
where $p$ is the pressure, $\rho$ and $\varepsilon$ are the rest mass
density and the total energy density respectively, and $K$ and
$\Gamma$ are the polytropic parameters.  For a central density $\rho_c
= 5.87 \times 10^{-4}~\rm{km}^{-2}$ and polytropic parameters $K =
217.858~\rm{km}^2$ and $\Gamma = 2$, the TOV equations provide a
stellar model with typical mass $M = 1.40 M_{\odot}$ and radius $R =
14.151~\rm{km}$~\cite{Dimmelmeier:2005zk}.

%
\section{Perturbative Framework} 
\label{Sec:framework}
%

The system of perturbation equations for studying the coupling between
the radial and nonradial oscillations of spherical star has been
introduced in~\cite{Passamonti:2004je, passamonti-2006,
Passamonti:2005cz}. Therefore in this paper we only report the main
properties of the perturbation framework and equations. For more
details see the references cited above.

The study of linear perturbations of a spherically symmetric
background can be significantly simplified by adopting the expansion
of the perturbed quantities in scalar, vector and tensor
harmonics. This technique enables us to separate in the perturbed
quantities the angular dependence from the time and radial parts. This
reduces the problem to just one spatial dimension. Any harmonic
component of the linear perturbative variables is then identified by
the harmonic indices $(\ell,m)$ and is dynamically independent.  A
further sub-classification can be carried out on a spherically
symmetric background, where two perturbation classes with opposite
parity and independent dynamics can be defined, namely the axial
(\emph{odd-parity}) and the polar (\emph{even-parity}) perturbations.

The above perturbative technique can be extended to second order
perturbations. However, due the hierarchical structure of the
perturbation methods the nonlinear perturbations obey inhomogeneous
perturbation equations, where the homogeneous part has the same
differential structure as the linear perturbation equations, whereas
the source terms are made by the product of linear perturbations. From
the form of these quadratic terms, we can notice that the $(\ell,m)$
second order perturbations depends on self coupling terms, which
couple linear perturbations with the same harmonic index $\ell$, and
mixed terms, which are instead the product of linear perturbations
with different harmonic indices. 

In this paper as well as in~\cite{Passamonti:2004je, passamonti-2006,
Passamonti:2005cz}, we select the coupling between the radial $(\ell=0)$
and nonradial $(\ell \ge 2)$ perturbations, and concentrate on the
numerical simulations of the polar perturbative class. For the
coupling between the radial and axial nonradial oscillations
see~\cite{passamonti-2006, Passamonti:2005cz}.
In order to study the gauge properties of the linear and nonlinear
perturbations, we have found convenient to use the 2-parameter
relativistic perturbation theory~\cite{Bruni:2002sm} and the gauge
invariant formalism of Gerlach and Sengupta~\cite{Gerlach:1979rw,
Gerlach:1980tx, Gundlach:1999bt, Martin-Garcia:2000ze}, in the version
developed by~\cite{Gundlach:1999bt, Martin-Garcia:2000ze}, that we
call GSGM formalism.
We have labelled the radial and nonradial perturbation with two
distinct parameters and studied the gauge invariance of our nonlinear
quantities. This particular coupling can then be studied by means of
variables that are gauge invariant when the linear radial perturbation gauge is
fixed. Adopting the so-called {\it radial gauge} for the radial
perturbations, the problem becomes essentially that of linear polar
perturbations on a time-dependent spherical background.

The remainder of this section is dedicated to the description of the
main properties of the radial and  nonradial perturbations and 
their coupling terms. We label linear radial quantities by
a ${}^{(1,0)}$ superscript, linear nonradial quantities by a 
 ${}^{(0,1)}$ superscript and coupling terms by a ${}^{(1,1)}$ 
superscript.
 The radial pulsations are described by a set of four perturbative
fields, namely two metric quantities~$S^{(1,0)}$, $\eta^{(1,0)}$ and
two fluid variables, which are the enthalpy~$H^{(1,0)}$ and the
velocity~$\gamma^{(1,0)}$ variables. They obey three first order in
time evolution equations and two constraints, as there is a single
radial degree of freedom. We can then solve a hyperbolic-elliptic
system of equations, which is formed by two hyperbolic equations,
for~$\gamma^{(1,0)}$ and~$H^{(1,0)}$, and the Hamiltonian constraint
that at any time step updates the variable~$S^{(1,0)}$,
while~$\eta^{(1,0)}$ is determined by the second elliptic equation.

The linear polar nonradial oscillations can be studied with a set of
gauge invariant quantities~\cite{Gundlach:1999bt,
Martin-Garcia:2000ze, Passamonti:2004je}, the three metric
perturbations~$S^{(0,1)}$, $k^{(0,1)}$, $\psi^{(0,1)}$ and the three
fluid variables~$H^{(0,1)}$, $\gamma^{(0,1)}$, $\alpha^{(0,1)}$.
Among the different systems of perturbation equations that are
available in the literature, we choose a system of three partial
differential equations for the three variables~$S^{(0,1)}$,
$k^{(0,1)}$ and $H^{(0,1)}$. It consists of two hyperbolic equations,
which describe respectively the gravitational waves and sound waves
propagations, and an elliptic equation, i.e the Hamiltonian
constraint~\cite{Nagar:2004ns, Nagar:2004pr, Passamonti:2005cz}.
Since the Hamiltonian constraint is used for updating at any time step
one of the unknowns of the problem, the errors associated with the
violation of this constraint are corrected.
The other nonradial variables $\psi^{(0,1)}$, $\gamma^{(0,1)}$,
$\alpha^{(0,1)}$, which are necessary for updating the source terms of
the second order perturbative equations, can be obtained by three
hyperbolic partial differential equations that are reported in
Appendix~\ref{sec:non-radial-eqs}.

The nonlinear coupling terms obey the same system of equations as
the linear nonradial terms, but with nonvanishing source terms 
in the region interior to the star.
Therefore, we evolve the two metric quantities $S^{(1,1)}$ and
$k^{(1,1)}$ and the fluid variable~$H^{(1,1)}$. 
In order to understand the structure of the perturbative framework we
define the following set of linear and nonlinear perturbations: 
\begin{align}
& \mathcal{R}^{(1,0)} \equiv \left( S^{(1,0)}, \eta^{(1,0)}, H^{(1,0)},
                        \gamma^{(1,0)} \right) \, ,  \label{R10} \\ 
& \mathcal{N}^{(0,1)} \equiv \left( S^{(0,1)}, k^{(0,1)}, 
                        \psi^{(0,1)}, H^{(0,1)}, \gamma^{(0,1)},
                        \alpha^{(0,1)} \right) \, , \label{N01} \\ 
& \mathcal{Q}^{(1,1)} \equiv \left( S^{(1,1)}, k^{(1,1)}, H^{(1,1)} \right) 
\, , \label{Q11}
\end{align}
and write the system of the three perturbative equations for the
coupling as follows:
\begin{equation}
\mb{L}_{NR} \left[ \mathcal{Q}^{(1,1)} \right] = \mb{S}
                    \left[ \mathcal{R}^{(1,0)}  \otimes \mathcal{N}^{(0,1)} 
\right]
                    \ , 
\label{eqs-coupl-scheme}
\end{equation}
where $\mb{L}_{NR}$ is an operator representing the homogeneous part
of the system, which has the same differential structure as the
linear nonradial perturbation equations for the variables $S^{(1,1)}$,
$k^{(1,1)}$ and $H^{(1,1)}$.  The source terms are instead represented
by the operator $\mb{S}$, which also contains spatial
and first order time derivatives.

Outside the star, the linear polar nonradial perturbations have just 
one degree of freedom, which is described by means of the Zerilli-Moncrief
function~\cite{Zerilli:1970fj}. In terms of the gauge-invariant 
variables it is is given by~\footnote{In the
reference~\cite{Passamonti:2004je} there is a typo in the definition
of the Zerilli-Moncrief function that here we have corrected. In the same
paper, equation~(79) must be corrected as follows: $U=u^Av_A=\lambda
\frac{e^{-\Lambda}}{r}\gamma^{(1,0)}$.}
\begin{align}
 Z^{(0,1)} & =  \frac{4 r^2 e^{-2\Lambda}}{ \ell \left(\ell+1\right) \left[(\ell(\ell+1)-2)r+6M \right]} 
                 \Bigg[r S^{(0,1)} \nn \\
           & +    \frac{1}{2}\left(\ell\left(\ell+1\right) +\frac{2M}{r}\right)   
             e^{2\Lambda} k^{(0,1)} - r k_{,r}^{(0,1)}  \Bigg] \ ,  
\label{Zer-def}
\end{align}
where $M$ is the mass of the star. The radial-nonradial coupling terms have 
the same angular dependence and satisfy an equivalent Zerilli-Moncrief
wave-like equation as the linear nonradial perturbation. We denote this 
variable as $Z^{(1,1)}$; it is constructed from an expression
of the form of Eq.~(\ref{Zer-def}) with all the variables of type
$^{(0,1)}$ replaced by the corresponding $^{(1,1)}$ ones. For any 
multipole $(\ell,m)$, the total Zerilli-Moncrief function is obtained as
\begin{equation}
\label{eq:ztotal}
Z_{\ell m}\equiv Z \equiv Z^{(0,1)}+Z^{(1,1)} \ ,
\end{equation}
where we have explicitly reintroduced the multipolar indices. As a result, 
the total emitted power in gravitational radiation (at infinity) is 
computed as (see for example~\cite{2006CQGra..23.4297N})
\begin{equation}
\frac{d E}{d t} = \frac{1}{64 \pi} \sum_{\ell\ge 2, m} \, \frac{\left( \ell + 2
\right)\, !}{\left(\ell-2\right)\, !} \, |\dot{Z}_{\ell m}|^2 \ ,
\label{GWPower}
\end{equation}
where the overdot denotes time derivative.

%
\section{Setting up Numerical Simulations}
\label{Sec:numcode}
%

The numerical code for studying the dynamical evolution of polar
coupling perturbations has the same structure as the code developed
in~\cite{passamonti-2006, Passamonti:2005cz} for the axial coupling
case. We had to replace the parts of the code that compute the axial
linear and coupling perturbations with the new routines that are
appropriate for the polar sector. For details about the overall
structure of the code see~\cite{Passamonti:2005cz, passamonti-2006},
while for the part that solves the polar nonradial perturbations
see~\cite{Nagar:2004ns, Nagar:2004pr}.  This latter part has also been
used for treating the equations for the radial-nonradial terms, by
adding the source terms found in~\cite{Passamonti:2004je,
passamonti-2006}.

\subsection{Initial Data \label{sec:id}}

The independence of linear perturbations from the harmonic index
$\ell$ implies that we need to separately excite the radial and
nonradial oscillations. In order to simplify the study of numerical
simulations and identify the correct nonlinear harmonics we have
excited the linear perturbations by selecting specific modes.  For
radial pulsations we have set up a Sturm-Liouville problem for the
radial velocity perturbations and solved it numerically with
relaxation methods. As tested
in~\cite{Passamonti:2005cz,passamonti-2006}, this code determines the
eigenfrequencies of radial modes with an accuracy to better than
$0.2\%$ with respect to the published values.  With an appropriate
choice of the initial phases for the various perturbed variables, we
can excite the radial eigenmodes by providing only the eigenfunctions
associated with the radial velocity perturbation~$\gamma^{(1,0)}$.
The simulations are numerically stable over a large multiple of the
largest oscillation period that we are interested in.  Using Fast
Fourier Transformations (FFT) of the time evolution of selected
variables, we can reproduce the linear mode frequencies of radial
modes with an accuracy to better than $0.4\%$, as shown in
Table~\ref{tab:mode-freq}.  More details can be found
in~\cite{Passamonti:2005cz, passamonti-2006}.

There are various sets of initial conditions for exciting the polar
nonradial perturbations which satisfy the Hamiltonian and momentum
constraints on the initial slice~\cite{Allen:1998xj, Ruoff:2001ux,
Nagar:2004ns}. After having explored all the different choices
mentioned in the literature, we have not noticed any important
qualitative difference with respect to the coupling
perturbations. Therefore, in this paper we consider only a
representative case, which consists in perturbing the fluid with an
enthalpy eigenfunctions, imposing $S^{(0,1)}=0$ (conformally flat
initial data) and determining $k^{(0,1)}$ by solving the Hamiltonian
constraint.

\begin{table}[!t]
\caption{\label{tab:mode-freq} Eigenfrequencies (in kHz) of the first
  eight fluid modes of the radial and nonradial $\ell=2$ polar perturbations.
  The second column corresponds to the results from calculation in the
  frequency domain (the solution of the Sturm-Liouville
  problem~\cite{Passamonti:2005cz}), while the third column displays
  the results of an FFT of the enthalpy time evolution. In square
  brackets we denote the relative difference between the frequencies
  extracted from the time evolution and the Sturm-Liouville
  solutions. The fourth column shows the frequencies of the polar
  nonradial oscillations, which are determined by an FFT of a
  time-domain simulation.}
\begin{center}  
\begin{ruledtabular}
\begin{tabular}{c | c | c | c | c }
  Radial     & Freq. code  & FFT   &  Nonradial   & FFT   \\ 
 \hline 
    F        & 1.443      & 1.437  [$0.4\%$]  &  ${}^2 \rm{f}$     & 
1.587   \\ 
 $\rm{H}_1$  & 3.955      & 3.951  [$0.1\%$]  &  ${}^2 \rm{p_1}$   & 
3.757   \\
 $\rm{H}_2$  & 5.916      & 5.913  [$0.05\%$] &  ${}^2 \rm{p_2}$   & 
5.699   \\
 $\rm{H}_3$  & 7.775      & 7.771  [$0.05\%$] &  ${}^2 \rm{p_3}$   & 
7.614   \\
 $\rm{H}_4$  & 9.590      & 9.587  [$0.03\%$] &  ${}^2 \rm{p_4}$   & 
9.419   \\
 $\rm{H}_5$  & 11.38      & 11.368 [$0.1\%$]  &  ${}^2 \rm{p_5}$   & 
11.25   \\
 $\rm{H}_6$  & 13.15      & 13.152 [$0.01\%$] &  ${}^2 \rm{p_6}$   & 
13.03   \\
 $\rm{H}_7$  & 14.92      & 14.916 [$0.02\%$] &  ${}^2 \rm{p_7}$   & 
14.781  \\
\end{tabular}
\end{ruledtabular}
\end{center}
\end{table}
\begin{figure}[!t]
\begin{center}
\includegraphics[width=85mm]{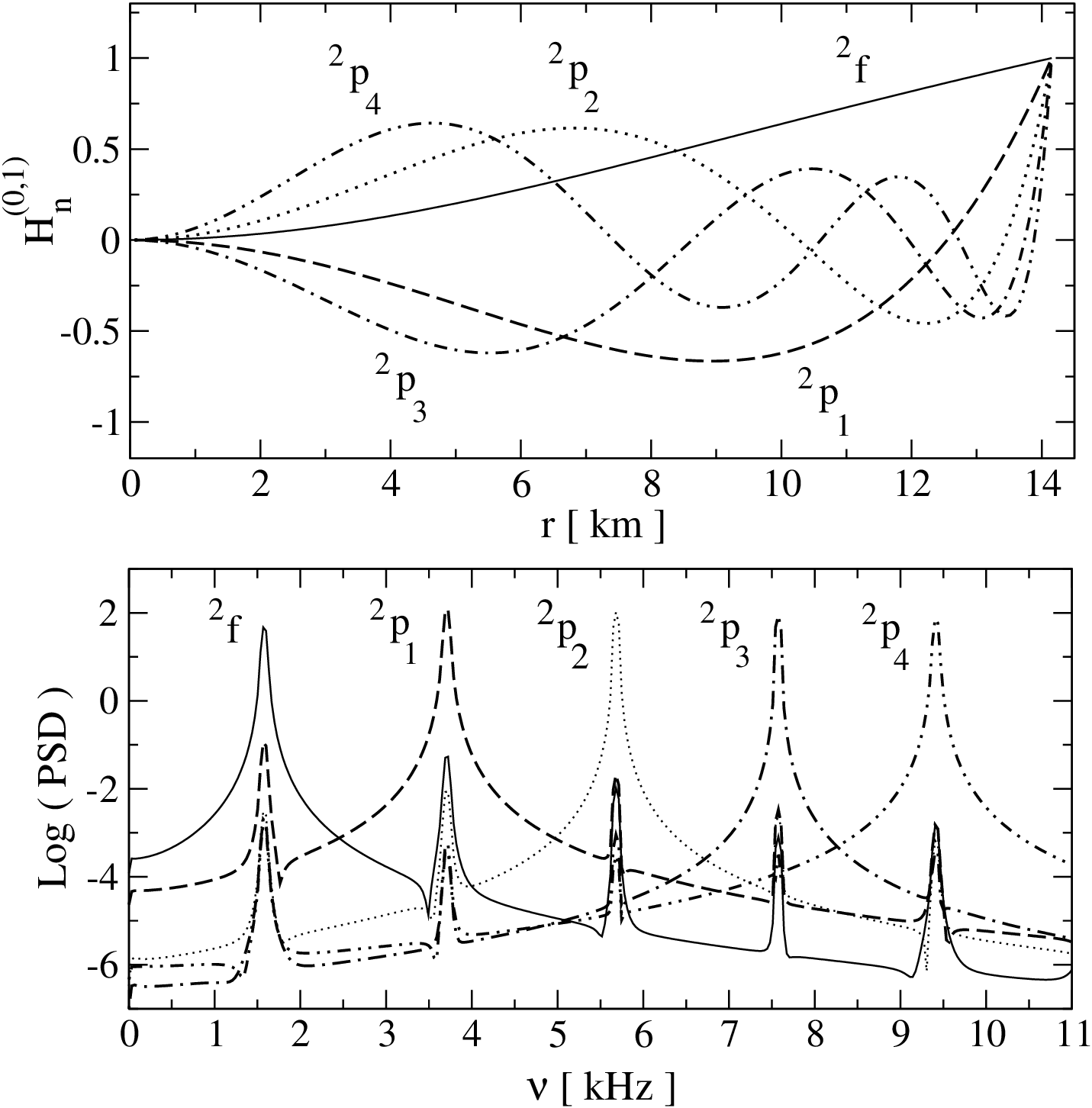} 
\caption{Normalized eigenfunctions of the perturbed enthalpy for the
fundamental nonradial mode ${}^2 \rm{f}$ and its first four overtones,
${}^2\rm{p}_1$ to ${}^2\rm{p}_4$ (\emph{upper panel}).  The 
\emph{lower panel}
displays the Fourier power spectral density of the evolution of the 
perturbed enthalpy, in simulations where each of the above eigenfunctions 
was used as an initial perturbation. In all cases, the desired mode is predominantly
excited, while other modes (still excited due to truncation errors)
have an amplitude which is several orders of magnitude smaller.
\label{fig:Heig_01}}
\end{center}
\end{figure}
The enthalpy eigenfunctions are obtained with the 
\emph{eigenfunction recycling} method developed 
in~\cite{Stergioulas:2003ep, Dimmelmeier:2005zk}, where the eigenfunction is extracted after a
simulation by means of FFT transformations at every grid point.
In a first simulation, we use trial eigenfunctions that excite various modes.
The eigenfunction of the desired mode is then extracted and used as
an initial perturbation for a second simulation, in which case the
desired mode is predominantly excited. This \emph{recycling}
procedure can be repeated until the amplitude of the other modes
becomes sufficiently small. Examples of extracted eigenfunctions are
shown in the top panel of Fig.~\ref{fig:Heig_01}, which displays
the eigenfunctions of the fundamental quadrupole mode ${}^2\rm{f}$ 
and its first four overtones ${}^2\rm{p}_1$ to ${}^2\rm{p}_4$, for 
the equilibrium model
mentioned in Sec.~\ref{Sec:equil}. The lower panel displays the Fourier
power spectral density (PSD)
of the time evolution of the perturbed enthalpy, in simulations where each
of the above eigenfunctions was used as an initial perturbation.  In
all cases, we observe that the desired mode is predominantly excited,
while other modes (still excited due to truncation errors) have an
amplitude which is several orders of magnitude smaller.
\begin{table}[!t]
\caption{Initial configurations (CASE I -- IV),
constructed by various linear combinations of eigenfunctions
corresponding to particular radial and nonradial modes. For the
relative amplitudes used in these linear combination, see the main 
text.}
\label{tab:IDconfig}
\begin{ruledtabular}
\begin{center}
\begin{tabular}{c | c | c  }
CASE &  Radial modes      &  Nonradial modes          \\
 \hline
    I          &   $\rm{F}$                &  ${}^2 \rm{f}$                 \\
    II         &   $\rm{F}$, $\rm{H}_1$, $\rm{H}_2$, $\rm{H}_3$, $\rm{H}_4$,
 $\rm{H}_5$   &  ${}^2 \rm{f}$                 \\
    III        &   $\rm{F}$                &  ${}^2 \rm{f} $, ${}^2
 \rm{p}_1$, ${}^2 \rm{p}_2$, ${}^2 \rm{p}_3$, ${}^2 \rm{p}_4$, ${}^2 \rm{p}_5$    \\
    IV         &   $\rm{F}$, $\rm{H}_1$, $\rm{H}_2$, $\rm{H}_3$, $\rm{H}_4$,
 $\rm{H}_5$   &  ${}^2 \rm{f} $, ${}^2 \rm{p}_1$, ${}^2 \rm{p}_2$, 
${}^2 \rm{p}_3$, ${}^2 \rm{p}_4$, ${}^2 \rm{p}_5$    \\
\end{tabular}
\end{center}
\end{ruledtabular}
\end{table}

\subsection{Case Studies \label{sec:case}}

The nonlinear coupling between radial and nonradial modes is studied
here in a number of particular cases, in which the perturbed initial
data consists of a linear combination of various radial and (quadrupole) 
nonradial modes. Table~\ref{tab:IDconfig} provides a summary of the 
modes used in each case.  In CASE I, only the fundamental radial and the
fundamental nonradial modes are present in the perturbed initial
data. CASE II includes, in addition, the first five radial overtones.
CASE III includes the fundamental radial and nonradial modes plus 
the first five nonradial overtones. Finally, CASE IV includes all 
radial and nonradial modes up to the fifth overtone.

The amplitude of each linear mode is determined by choosing a
particular value of its pulsational kinetic energy. For radial
pulsations, the kinetic energy is computed at an instant of
vanishing Lagrangian displacement (vanishing potential energy, 
maximum kinetic energy) as~\cite{Bardeen:1966tm}:
\begin{equation}
E_k^{(1,0)}  = 2 \pi \int_{0}^{R} dr 
            \left( \varepsilon + p \right) 
            \, e^{\Lambda + \Phi } 
            \left( r\, \gamma^{(1,0)} \right) ^2 \label{Kin_ener_10},
\end{equation}
where $\gamma^{(1,0)}$ is related to the perturbed radial velocity.

For nonradial pulsations, the kinetic energy can be determined in terms of
the GSGM quantities in the Regge-Wheeler gauge as follows~(see
Appendix~\ref{sec:Kin_Energy}):
\begin{align}
 E_k^{(0,1)}  = & \frac{1}{2} \int_{0}^{R} dr \left( \varepsilon + p
\right) e^{\Phi+\Lambda}  \nn \\
 &      \left[ r^2  \left( \gamma^{(0,1)} - \frac{\psi^{(0,1)}}{2}
        \right) ^2
          +  \ell  \left(\ell+1 \right) \left( \alpha^{(0,1)} 
       \right) ^2 \right] \label{Kin_ener_01} \, ,
\end{align}
where $\gamma^{(0,1)}$ and $\alpha^{(0,1)}$ are proportional to the
radial and longitudinal components of the velocity, respectively,
while the metric perturbation $\psi^{(0,1)}$ is proportional to
$\delta g_{r0}$.  In order to determine the amplitudes of the initial 
data, we average the pulsational kinetic energy of 
Eqs.~(\ref{Kin_ener_10})-(\ref{Kin_ener_01}) over several oscillation 
periods.

\begin{figure}[!t]
\begin{center}
\includegraphics[width=85mm]{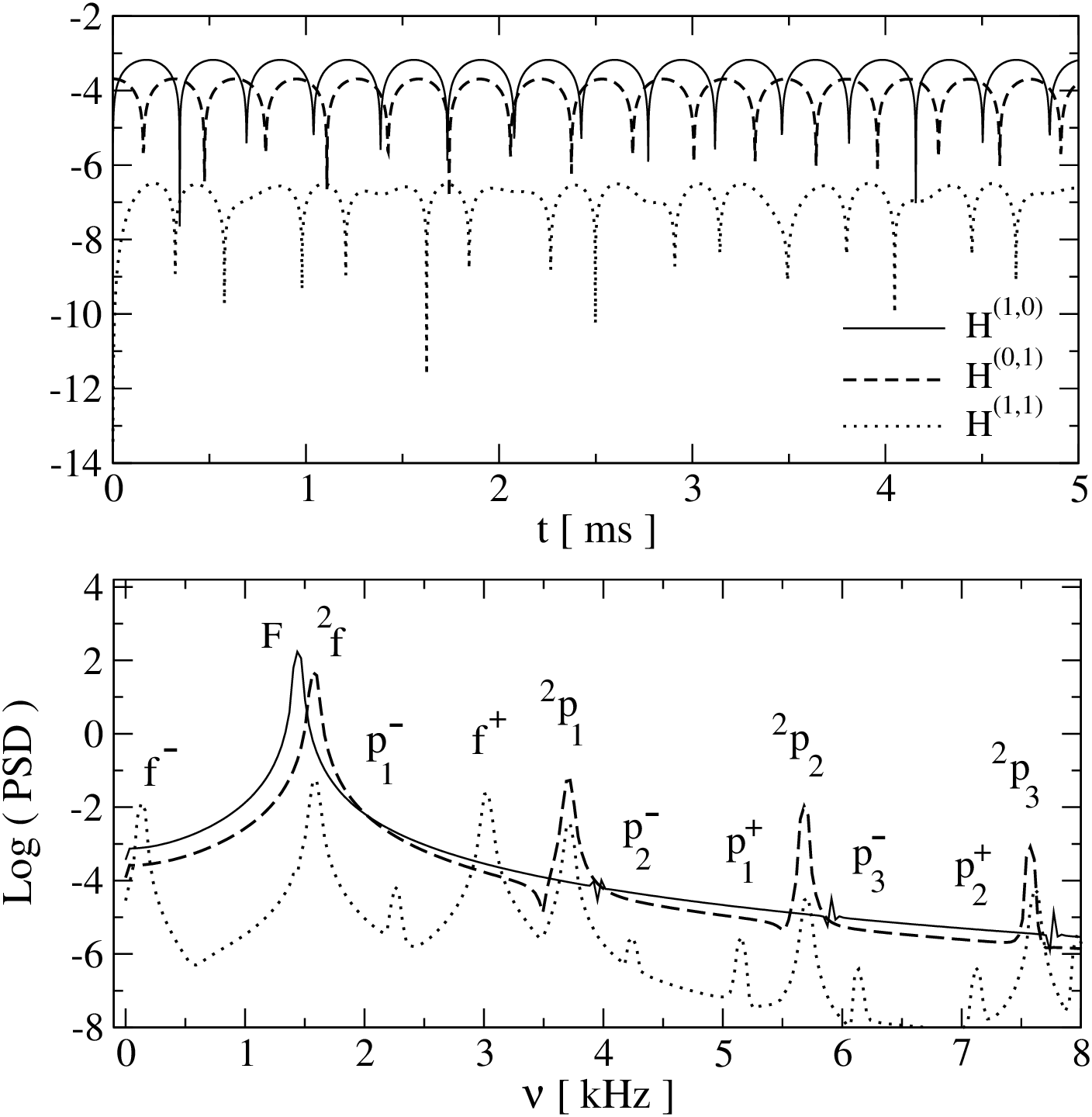}
\caption{Time evolution (\emph{upper panel}) and the spectral
properties (\emph{lower panel}) of the enthalpy for CASE I, when the
star is oscillating at first perturbative order in the radial and
nonradial fundamental modes.  In the (\emph{upper panel}), the
enthalpy is given as $\log|H^{(i,j)}|$, where $i,j=0,1$.  The radial
$H^{(1,0)}$ and nonradial $H^{(0,1)}$ enthalpy are shown with
\emph{solid} and \emph{dashed lines} respectively, while $H^{(1,1)}$
is represented with a \emph{dotted line}.  The nonlinear harmonics
have been denoted as $\rm{f}^{\pm} = {}^2 \rm{f} \pm \rm{F}$ and
$\rm{p}_{n}^{\pm} = {}^2 \rm{p}_{n} \pm \rm{F}$.}
\label{H-each-F-f2}
\end{center}
\end{figure}


\begin{figure}[!t]
\begin{center}
\includegraphics[width=85mm]{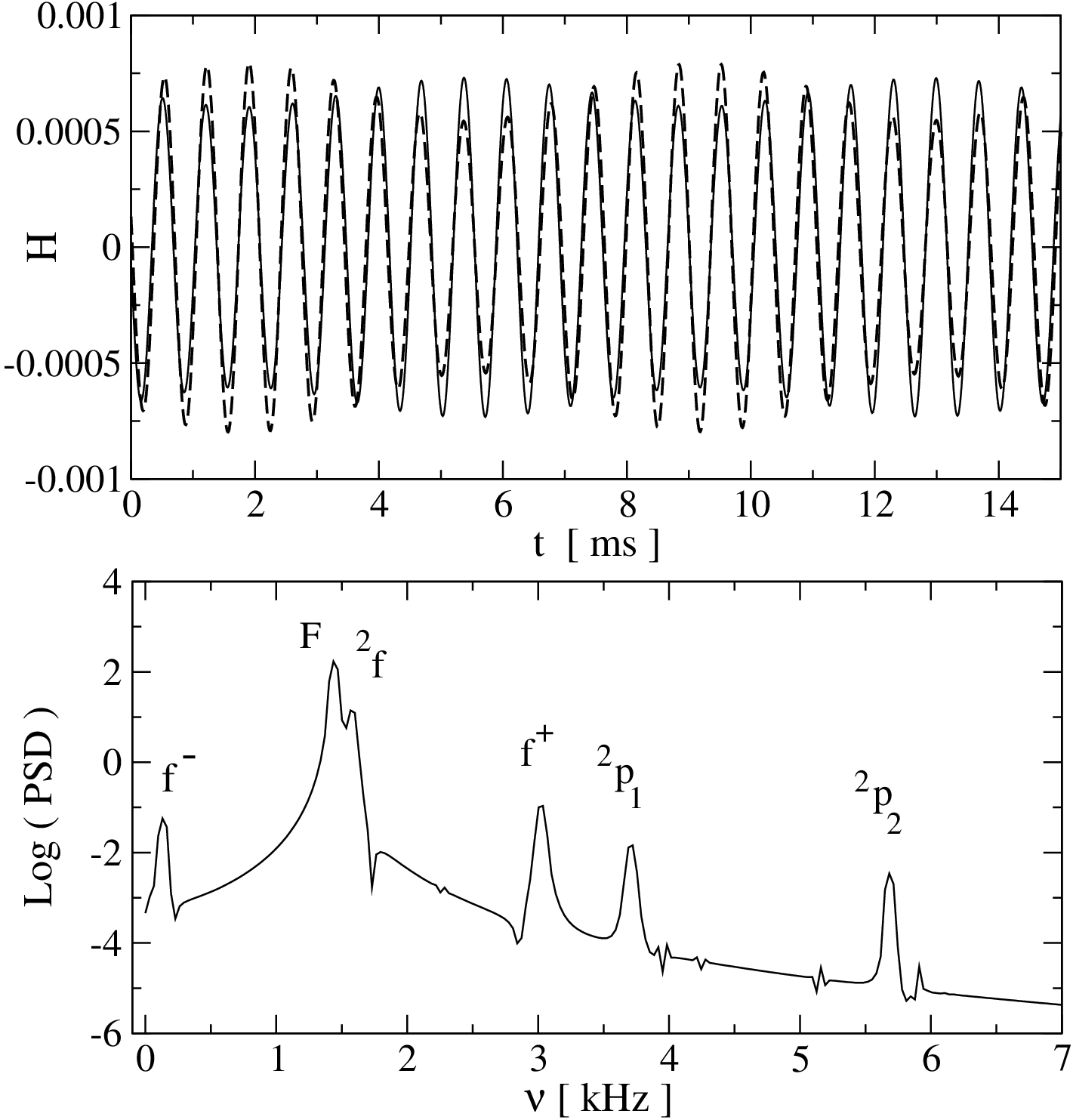}
\caption{For the same initial pulsating configuration of
Fig.~\ref{H-each-F-f2} (CASE I) the \emph{upper panel} depicts 
the total enthalpy $H$ as given by Eq.~(\ref{Htot}) in the equatorial 
plane ($\theta=\pi/2$, \emph{solid line}) and in the polar direction
($\theta=0$, {\it dashed line}). {\it Lower panel:} the Fourier 
power spectral density of $H$ in the equatorial plane.
\label{H-tot-F-f2}}
\end{center}
\end{figure}

After having determined the required amplitude for each mode to get a
certain pulsation energy, the radial pulsations are excited by a
linear combination of the variables $\gamma_n ^{(1,0)}$ (related to
the radial velocity eigenfunction), where the index $n$ denotes
different modes:
\begin{equation}
\gamma ^{(1,0)} = \sum _{n}  \gamma_{n} ^{(1,0)} \, . 
\end{equation}
Correspondingly, for nonradial oscillations, the linear combination 
of the enthalpy eigenfunctions $H^{(0,1)}_{n}$, used to initiate the
pulsations, is given by
\begin{equation}
H ^{(0,1)} = \sum _{n} H_{n} ^{(0,1)} \, . \label{H01-eig}
\end{equation}

For simplicity we consider vanishing initial data for all coupling
variables, i.e.  $\mathcal{Q}^{(1,1)} = 0$. The Hamiltonian and
momentum constraints are then violated on the initial Cauchy
hypersurface and the numerical solutions will be effected by an
initial transient of short duration~\cite{Passamonti:2005cz,
passamonti-2006}. This initial transient is not affecting the main
results presented here.

\subsection{Boundary Conditions}

To complete the description of the initial-value problem, one needs to
impose the boundary conditions at the origin, the stellar surface and
at infinity. This topic has been addressed in detail
in~\cite{Passamonti:2004je, Passamonti:2005cz, passamonti-2006} and
hence we do not repeat it here. However, it is worth mentioning that
the surface treatment used in the axial case~\cite{Passamonti:2005cz,
passamonti-2006} has been also implement for polar modes.  During the
evolution, some of the junction conditions for the coupling variables
could not be correctly imposed, due to the motion of the perturbed
stellar surface (see ~\cite{Sperhake:2001si, Passamonti:2005cz}).  We
address this issue as follows: i) at the stellar surface, we determine
the maximal Lagrangian displacement of the linear oscillations and ii)
we impose the junction conditions for the coupling variables on a
hypersurface which is always inside the oscillating star.  This method
is equivalent to removing a thin outer layer close to the stellar
surface, with a thickness depending on the pulsation amplitude. In
practice, for small-amplitude pulsations, only very few grid points
near the stellar surface of the equilibrium model need to be neglected
for the coupling variables. This does not have a noticeable
quantitative effect on our main results.

\begin{figure}[!t]
\begin{center}
\includegraphics[width=85mm]{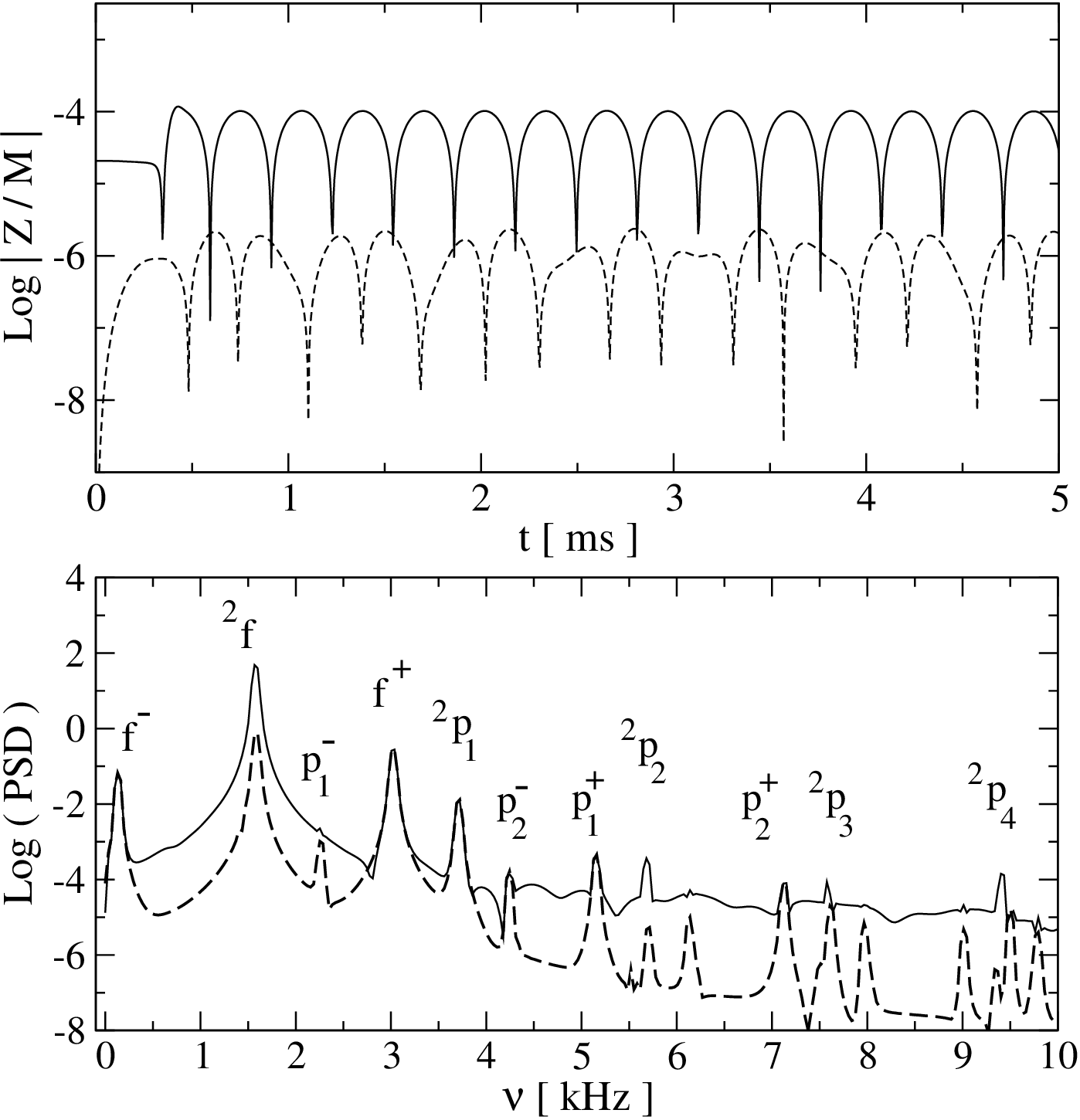}
\caption{Gauge-invariant waveforms ({\it top panel}) and Fourier power 
spectral density ({\it bottom panel}) for CASE I: we show together the coupling 
Zerilli-Moncrief function $Z^{(1,1)}$ ({\it dashed line}) and the total
one $Z\equiv Z^{(0,1)}+Z^{(1,1)}$ ({\it solid line}). 
\label{Zer-F-f2}}
\end{center}
\end{figure}

\section{Radial-Nonradial Mode Couplings}
\label{Sec:results}
%

We discuss now the numerical results for the coupling of
radial and nonradial pulsations, for the specific cases of initial
configurations described in Sec.~\ref{sec:case}.  In order to
interpret the numerical simulations, we first describe the main
properties that we expect from the structure of the equations 
for the coupling perturbative terms $\mathcal{Q}^{(1,1)}$.
The \emph{linear} radial oscillations will contain the
eigenfrequencies $\nu_i^{(1,0)}$ of the particular radial modes
excited by the initial data. Similarly, the \emph{linear} nonradial
oscillations, will contain the frequencies $\nu_i^{(0,1)}$ of the
particular nonradial modes.  The variables describing the
radial-nonradial coupling terms, $\mathcal{Q}^{(1,1)}$, which obey
equations~(\ref{eqs-coupl-scheme}), will then be driven by two
different kinds of frequencies: i) the natural frequencies
$\nu_n^{(0,1)}$, associated with the homogeneous part of the nonradial
equation~(\ref{eqs-coupl-scheme}), which are now excited by the
forcing term $\mb{S}$, and ii) the nonlinear coupling frequencies, the
so-called combination tones, whose frequencies are linear sums or
differences of linear mode frequencies.  For the second-order coupling
between radial and nonradial oscillations, an example of such a 
combination frequency is $\nu^{(1,1)\pm}=\nu^{(1,0)} \pm \nu^{(0,1)}$.

\begin{figure}[!t]
\begin{center}
\includegraphics[width=85mm]{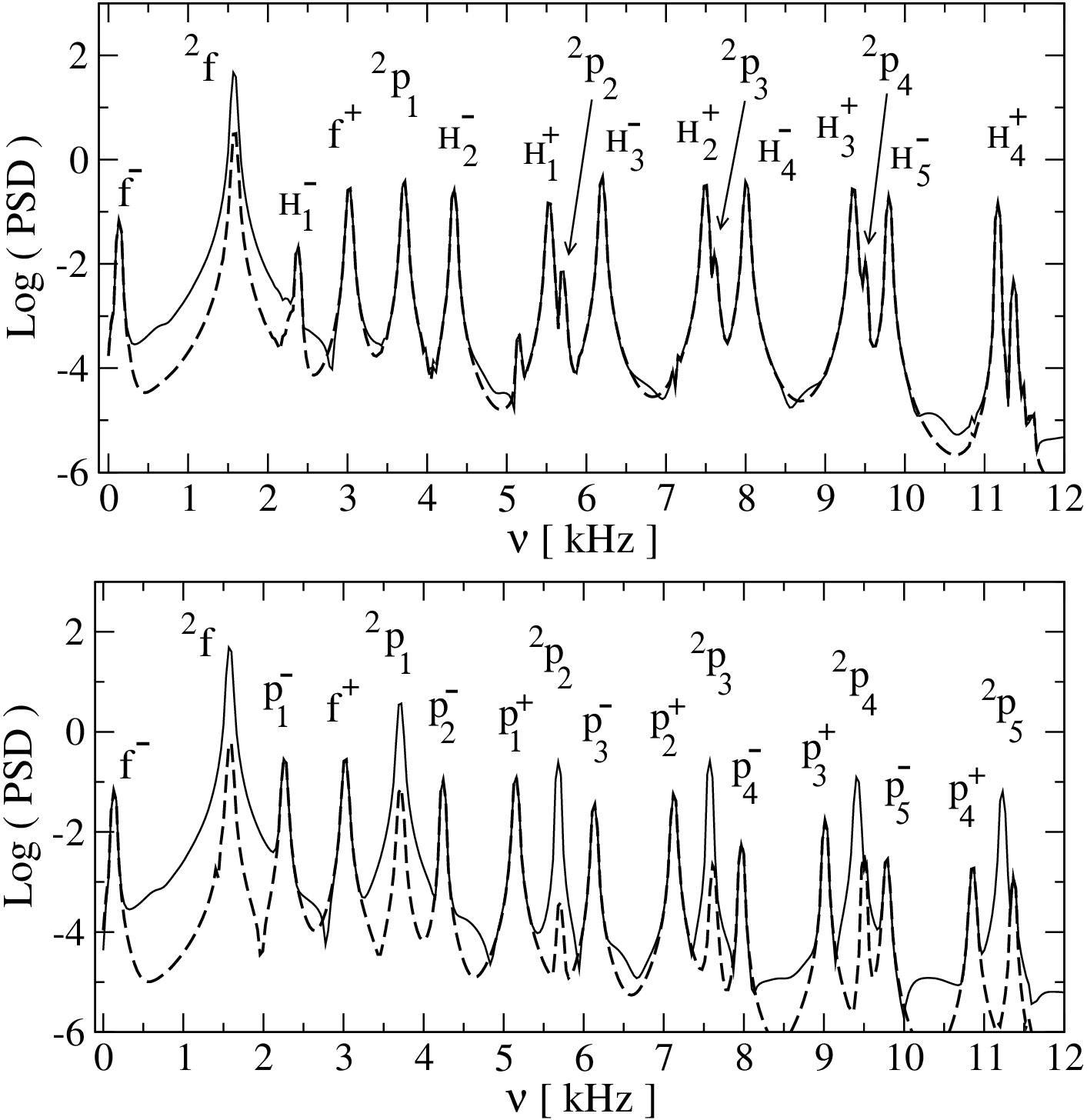}
\caption{Fourier power spectral density of $Z^{(1,1)}$ ({\it dashed
line}) and of $Z$ ({\it solid line}). The {\it top panel} refers to
CASE II and the {\it lower panel} to CASE III (see text for details).
In order to simplify the labelling of the figure we have defined
$\rm{H}_{n}^{\pm} = \rm{H}_{n} \pm {}^2 \rm{f}$, $\rm{f}^{\pm} = {}^2
\rm{f} \pm \rm{F}$ and $\rm{p}_{n}^{\pm} = {}^2 \rm{p}_{n} \pm
\rm{F}$.
\label{Zer-lincomb-A}}
\end{center}
\end{figure}

\subsection{Case I \label{sec:caseI}}

In CASE I, only the fundamental radial and the fundamental $\ell=2$
nonradial modes are present in the perturbed initial configuration. In
order to set the amplitude of each mode, we take into account recent
numerical simulations (see \cite{Dimmelmeier:2002bk,
Dimmelmeier:2002bm, Mueller2004, Ott2006, Dim2007} and references
therein) which suggest that over a total simulation time of about
20~ms the total energy emitted in gravitational waves can be of order
$10^{-8}~M_{\odot}$.  Furthermore, the post-bounce radial pulsations
in the above simulations induced a variation in the central energy
density of the order of 1-5\%.

In our numerical simulations, we find that a gravitational wave energy
of $1.11 \times 10^{-8}~M_{\odot}$ (as determined by time integration
of Eq.~(\ref{GWPower})) is emitted after 20~ms of continuous
monochromatic emission of the fundamental $\ell=2$ mode, when the latter
has an average kinetic energy of $\langle E_k^{(0,1)} \rangle = 5.0 \times
10^{-8}~M_{\odot}$. In practice, we stop our simulation at 34~ms, so
that the energy emitted in gravitational waves is $ 1.74 \times
10^{-8}~M_{\odot}$.  The radial pulsations are excited with an
amplitude that corresponds to an average kinetic energy $\langle E_k^{(1,0)} \rangle
= 5.0 \times 10^{-7}~M_{\odot}$, which leads to a variation in the
central energy density of 1\%.

The upper panel of Fig.~\ref{H-each-F-f2} displays the time evolution
of the enthalpy variables~$H^{(1,0)}$, $H^{(0,1)}$ and~$H^{(1,1)}$
averaged at any time step along the spatial coordinate. While the two
linear variables are nearly perfectly monochromatic, the bilinear term
$H^{(1,1)}$ contains several frequencies. In the PSD of $H^{(1,1)}$,
shown in the lower panel of Fig.~\ref{H-each-F-f2}, several
combination frequencies, such as ${\textrm f}^{\pm} = {}^2 \rm{f} \pm F$
and ${\rm p_n^{\pm} = {}^2 p_{n} \pm F}$, for $n=1,2,3$, can be
clearly identified. We have also verified that the amplitude of
$H^{(1,1)}$ is nearly the product of the two oscillation amplitudes
of~$H^{(1,0)}$ and~$H^{(0,1)}$. The top panel of Fig.~\ref{H-tot-F-f2}
shows the corresponding time evolution of the (averaged) total
enthalpy~$H$ (at $\theta=\pi/2$, solid line, and $\theta=0$, dashed
line) defined as
\begin{equation}
H  \equiv H^{(1,0)} +  \sqrt{\frac{5}{\pi}} \frac{3 \cos ^2 \theta - 1 }{4} 
                    \left( H^{(0,1)} + H^{(1,1)} \right) \  ,  
\label{Htot}
\end{equation}
where the normalization constant of the spherical harmonic $Y^{00}$
is contained in the radial variable~$H^{(1,0)}$. One can notice a
modulation of the amplitude, which is mainly due to the different
angular dependence of the radial and nonradial oscillations. 
The lower panel of Fig.~\ref{H-tot-F-f2} shows the PSD of~$H$ 
at $\theta=\pi/2$. Both linear radial and nonradial fundamental 
modes are present. Of the several combination frequencies present 
in $H^{(1,1)}$, only $\rm{f}^{\pm} = {}^2 \rm{f} \pm F$ have a large 
enough amplitude to be clearly visible here. We recall that the two 
lowest-order linear pressure mode frequencies seen in the lower panels 
of Fig.~\ref{H-each-F-f2} and Fig.~\ref{H-tot-F-f2} are present because 
the initial data for the linear fundamental radial mode are not perfect, 
but also contain additional frequencies with very small amplitudes.
The top panel of Fig.~\ref{Zer-F-f2} shows the time evolution of $Z^{(1,1)}$ (dashed line) 
and $Z$ (solid line) extracted at $r=100$ km. We note that the bilinear 
function $Z^{(1,1)}$ is about two order of magnitude smaller than $Z$. 
In the lower panel of the same figure, the corresponding PSD are displayed. 
The main combination frequencies are $\rm{f}^{\pm} = {}^2 \rm{f} \pm \rm{F}$,
whose detectability will be addressed in Sec.~\ref{Sec:Strain} below.

\subsection{Cases II and III }
\label{sec:caseIIandIII}

Next, we study two initially pulsating configurations where several
modes are contemporarily excited. In CASE II, we excite the nonradial 
fundamental mode ${}^2{\textrm f}$ and a linear combination of radial 
modes up to the fifth overtone. In CASE III, we excite the radial 
fundamental mode ${\textrm F}$ and a up to the fifth pressure nonradial mode.
The pulsation amplitudes of the two fundamental modes $\textrm F$ and
${\rm {}^2f}$ have the same value as in CASE I. For the overtones, we
assume that the kinetic energy decreases proportionally with the order
of the mode. This assumption is motivated by the spectral properties
of the gravitational waves emitted in core collapse simulation. In
particular, we demand that each overtone or pressure mode of order $n$
has one fifth of the energy stored in the $n-1$ mode; i.e.,
\begin{eqnarray}
\langle E_k^{(1,0)} \rangle _{\rm{H}_n}  & = & \frac{1}{5}  
                            \langle E_k^{(1,0)} \rangle  _{\rm{H}_{n-1}}  
              \label{over-E} \, , \\
\langle E_k^{(0,1)} \rangle _{{}^2p_{n}} & = & \frac{1}{5} 
                            \langle E_k^{(0,1)} \rangle  _{{}^2p_{n-1}}   
              \label{pres-E} \, , 
\end{eqnarray}
where for $n=0$ the $\rm{F}$ and ${}^2\rm{f}$ modes are implied.
Notice that due to phase-cancellations, the total energy, when
exciting several modes, is not a linear sum of the individual
energies. Using Eq.~(\ref{Kin_ener_10}) and Eq.~(\ref{Kin_ener_01})
for the total perturbation, we determined that when the radial or
nonradial oscillations are excited up to the fifth overtone, the
radial and nonradial average kinetic energies are $\langle E_k^{(1,0)}\rangle =
6.25 \times 10^{-7}~M_{\odot}$ and $\langle E_k^{(0,1)}\rangle = 6.0 \times
10^{-8}~M_{\odot}$ respectively.

For CASE II, the upper panel of Fig.~\ref{Zer-lincomb-A} displays 
the PSD of the bilinear Zerilli-Moncrief function $Z^{(1,1)}$ as well 
as that of the total $Z$. On the other hand, the bottom panel of 
the same figures exhibits the same information for CASE III.
The spectra clearly show the presence of several expected combination 
frequencies: in particular, for CASE III, a repetitive triplet structure 
(such as, e.g. ${\rm p_1^+, p_2, p_3^-}$) is evident in the plot. 

Although we have chosen to show only a few representative cases, 
the bilinearity of all coupling variables is well reproduced 
by our numerical results. It is therefore trivial to scale our
results to any other desirable values of the radial and nonradial
average kinetic energies.

%
\section{Gravitational Waves}
\label{Sec:Strain}
%

\begin{figure}[!t]
\begin{center}
\includegraphics[width=85mm]{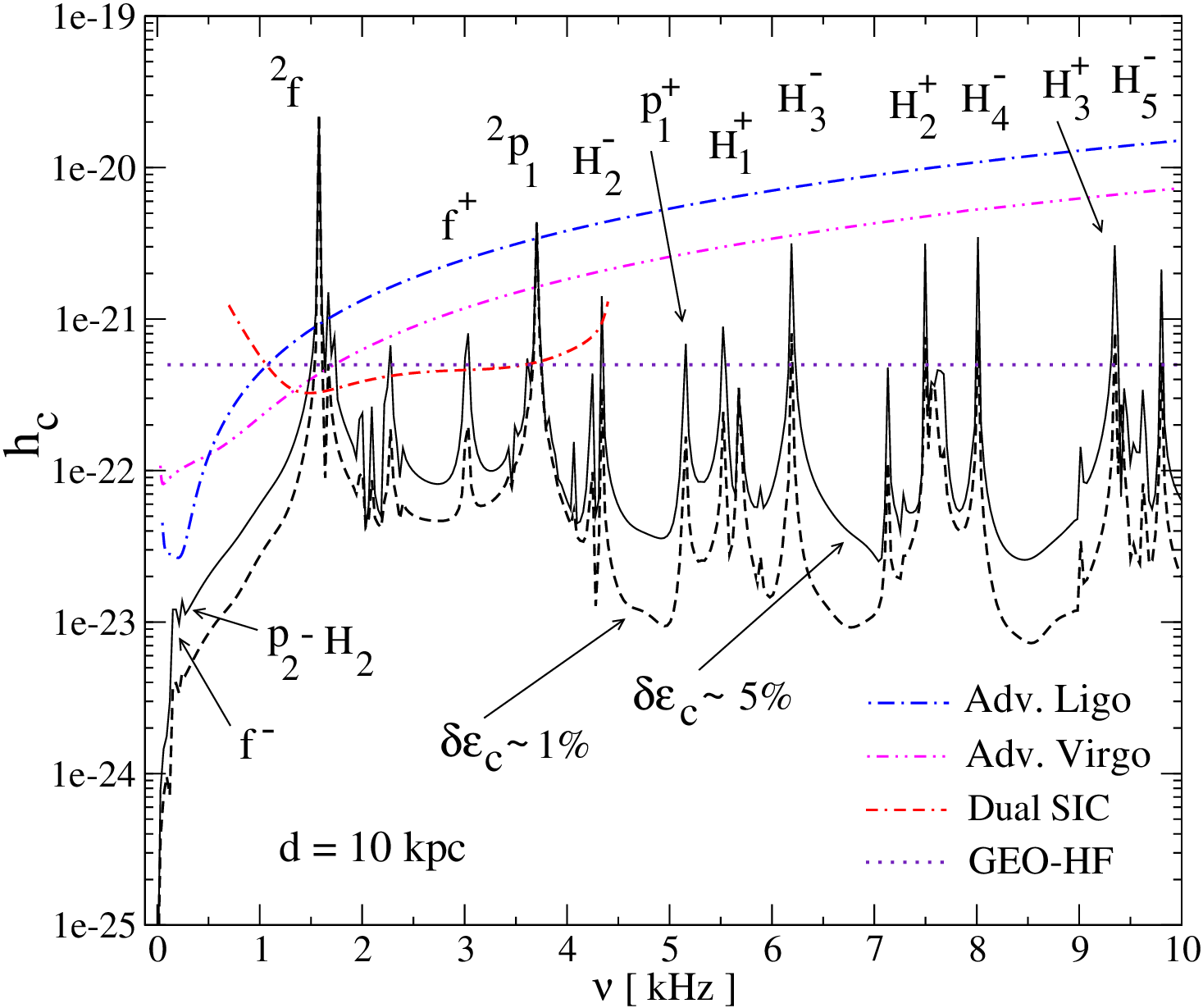} 
\caption{For a galactic source ($d=10$~kpc), the figure displays the
  characteristic strain $h_c(\nu)$ of two initial pulsating configurations
  for CASE IV of Table~\ref{tab:IDconfig}.  In the
  first~(\emph{dashed line}) and second~(\emph{solid line})
  configuration, the radial pulsations respectively lead to $1\%$ and
  $5\%$ variation of the central energy density. The figure also shows
  the $h_{\rm rms}(\nu)$ strain sensitivity, Eq.~(\ref{eq:hrms}), 
  of AdvLIGO, AdvVIRGO, Dual SIC and GEO-HF detectors (see legend). 
  Some of the linear and bilinear harmonics are within the sensitivity 
  window of these detectors, especially in the 
  $\delta \varepsilon_c \sim 5 \%$ case. 
\label{Strain-all}}
\end{center}
\end{figure}

In order to assess the detectability of gravitational waves emitted at
combination frequencies, when several oscillation modes are present,
we focus on CASE IV of Table~\ref{tab:IDconfig}, where the fundamental
radial and nonradial modes are present together with their first five
overtones. The average kinetic energy in the fundamental modes is
given as in Sec.~\ref{sec:caseIIandIII} with the energy in the
overtones given again by Eq.~(\ref{over-E}) and Eq.~(\ref{pres-E}).
We also vary the average kinetic energy in the radial modes, between a
minimum value $\langle E_k^{(1,0)}\rangle = 6.25 \times
10^{-7}~M_{\odot}$, that results in a central energy density variation
of 1\%, and a maximum value $\langle E_k^{(1,0)}\rangle = 9.96 \times
10^{-6}~M_{\odot}$, that results in a central energy density variation
of 5\%.  We compute the {\it characteristic} strain of gravitational
waves, defined as in~\cite{1998PhRvD..57.4535F}
\begin{equation}
 h_c \left( \nu \right) \equiv \dfrac{\sqrt{2}}{\pi d } 
\sqrt{ \dfrac{d E}{d \nu} } \, , 
\label{eq:h}
\end{equation}
where $ d E / d \nu $ is the energy spectrum of the gravitational
signal and $d$ is the distance of the source. The energy spectrum
can be written as~\cite{2006CQGra..23.4297N}
\begin{equation}
 \frac{d E}{d \nu}    =  \dfrac{\pi}{8} \, \sum_{\ell\geq 2}
                         \dfrac{\left(\ell+2\right)!}{\left(\ell-2\right)!} \,  
                          \nu ^2  \, |\hat{Z}_{\ell m}| ^2  \, , 
\label{eq:energy-spect}
\end{equation}
where $\hat Z_{\ell m}$ denotes the Fourier transform of the (total) 
Zerilli-Moncrief  function given by Eq.~(\ref{eq:ztotal}). 
For $\ell=2$, we have
\begin{equation}
 h_c \left( \nu \right) = \sqrt{\frac{6}{\pi}} \  \dfrac{\nu}{d} \, 
                  |\hat{Z}_{20}| \, .
\label{eq:h-Zer}
\end{equation}
Figure~\ref{Strain-all} shows $h_c(\nu)$ for a source at the galactic
distance $d=10$~kpc with an initial oscillating configuration given by
CASE IV and for an emission time of 34~ms. On top of this,
Fig.~\ref{Strain-all} also displays the optimal rms noise, $h_{\rm
rms}(\nu)$ for several planned detectors: the interferometric advanced
LIGO~\cite{ligo}, the advanced
VIRGO~\cite{WP-Adv-VIRGO,Ad_VIRGO_site}, the
GEO-HF~\cite{Willke:2006uw} designs and the proposed wide-band dual
resonant detector Dual SIC~\cite{2003PhRvD..68j2004B}.  We recall that
the optimal rms noise is defined as
\begin{equation}
h_{\rm rms} \left( \nu \right)   \equiv  \sqrt{ S_h \, \nu} \, ,
\label{eq:hrms}
\end{equation}
where $S_h(\nu)$ is the spectral density of the detector strain noise.
For signals with random direction and random orientation with respect
to the detector, the rms noise should be increased by a factor of 
$\sqrt 5$ \cite{1998PhRvD..57.4535F}.

Of the several linear nonradial modes excited in the simulation for
CASE IV, only the two lowest-order ones, ${}^2\rm{f}$ and ${}^2p_1$ have a
characteristic amplitude above the rms noise of either Advanced
LIGO (AdvLIGO) or Advanced VIRGO (AdvVIRGO). When a 1\%
central density variation for the radial oscillations is assumed, then
all bilinear combination frequencies are also below the strain sensitivity,
 for the above two detectors. However, when a 5\% central
density variation for the radial oscillations is assumed, then some
bilinear combination frequencies are above the rms noise of either
AdvVIRGO or both AdvLIGO and AdvVIRGO.

For the GEO-HF detector, a large number of bilinear harmonics is above
the strain sensitivity curve at a 5\% central density variation of
radial oscillations. With such a detector, even harmonics of up to 
10 kHz could become detectable. At the lower limit of a 1\%
central density variation, some bilinear harmonics have a characteristic
amplitude which is marginally above the strain sensitivity curve.

\begin{figure}[!t]
\begin{center}
\includegraphics[width=85mm]{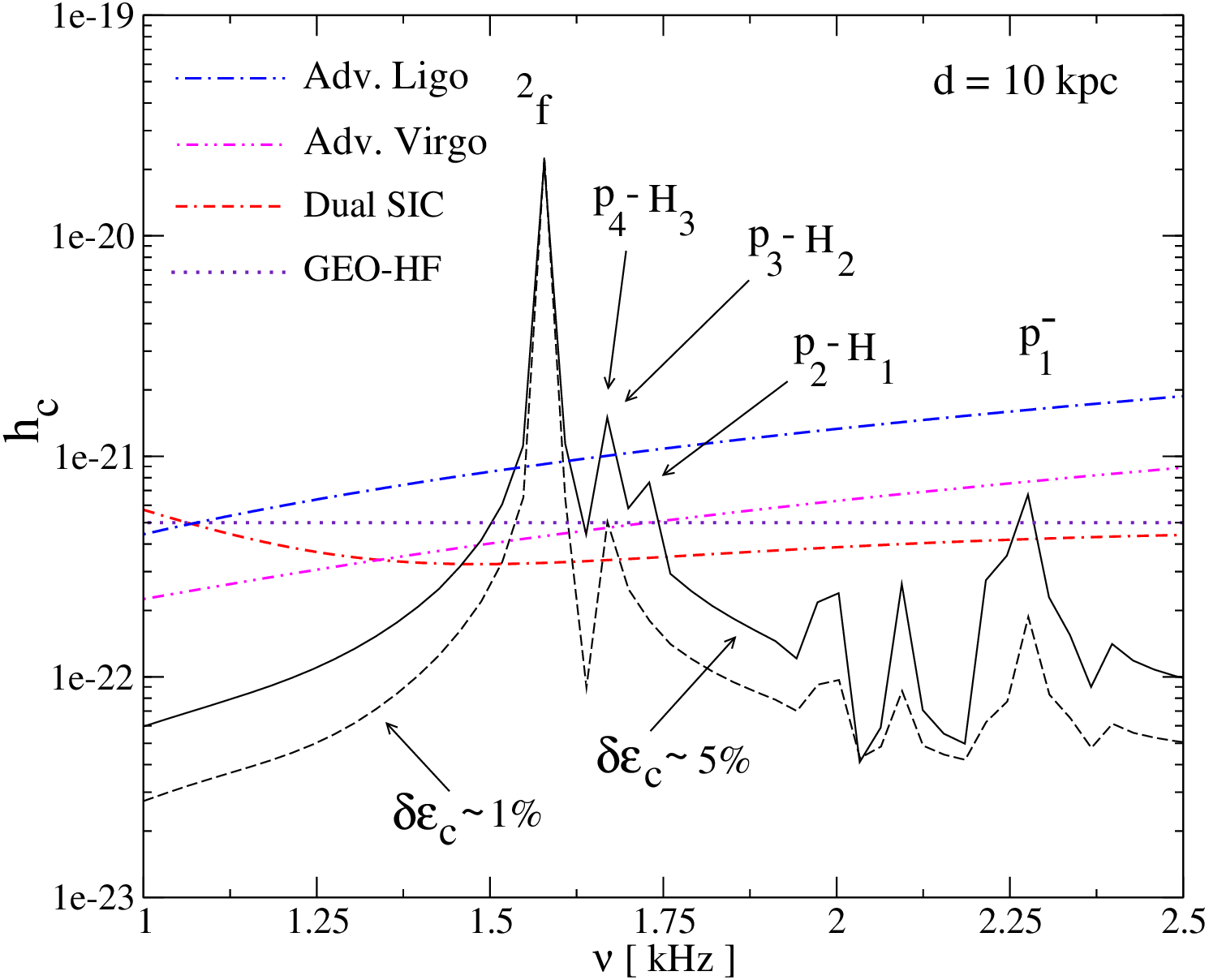}
\caption{Blow up of Fig.~\ref{Strain-all} in the region where the chance 
of a detection if higher. Besides the linear ${\rm {}^2 f}$ and 
${\rm {}^2 p_{1}}$ modes, there are several combination tones with a 
characteristic strain larger than the detector noise, in particular 
for the $\delta\varepsilon_c \sim 5 \%$ configuration. 
Note that the combination tones ${\rm \rm{p}_3 - \rm{H}_2}$ and 
${\rm \rm{p}_{4}-\rm{H}_3}$ fall into the same frequency bin and then
are indistinguishable.
\label{strain-high-band}}
\end{center}
\end{figure}

Figure~\ref{strain-high-band} focuses on the 1 to 2.5 kHz region of
Fig.~\ref{Strain-all}. The detectable bilinear combination
frequencies are close to the frequency of the fundamental ${\rm {}^2f} $
mode. Specifically, the ${\rm {}^2p_3-H_2}$ and the ${\rm {}^2p_4-H_3}$
combination frequencies fall into the same frequency bin 
and their combined amplitude exceeds the noise curve for both the
AdvLIGO and AdvVIRGO detectors, for a 5\% central density variation.
It is thus apparent, that some bilinear combination frequencies may be
interesting even for the AdvLIGO or AdvVIRGO designs.  For the Dual
SIC detector design, several bilinear harmonics fall within its
sensitivity window in the range of up to 4.5~kHz.
Figure~\ref{strain-high-band} shows that the ${\rm {}^2p_4-H_3}$
harmonic could become detectable even at a 1\% central density
variation.  Higher signal-to-noise ratios and better event rates can
be achieved with more advanced designs, such as the design for the
EURO detector in xylophone mode~\cite{euro}.

%
\section{DISCUSSION}
\label{Sec:discussion}

Using a gauge-invariant perturbative formalism, we studied the
bilinear coupling between the radial and nonradial polar pulsations in
a relativistic star. For typical values of mode-energies expected in
the post-bounce phase of core-collapse supernovae, we find that some
bilinear combination frequencies may become detectable with planned
interferometric detectors, such as Advanced VIRGO and GEO-HF, or with
proposed wide-band resonant detectors, such as the Dual SIC.

The possible detection of bilinear combination frequencies, in
addition to the expected linear mode frequencies, would yield
significant new constraints for the high-density equation of state of
compact stars, as they also contain information on the radial modes of
the star.  Using empirical relations constructed for
gravitational-wave
asteroseismology~\cite{Andersson:1996ak,2001MNRAS.320..307K,benhar-2004-70}, 
in combination with numerical data for radial modes in realistic 
compact-star models~\cite{Kokkotas:2000up} could allow for the 
determination of both the mass and radius of the star with good accuracy.

Bilinear combination frequencies of rotating proto-neutron stars were
recently studied in~\cite{Dimmelmeier:2005zk} where it was found that
rotational changes result in the difference in frequency between two
overtones coming close to the frequency of a fundamental mode. For
such cases it was suggested that these crossings could result in an
enhanced gravitational-wave emission at the combination frequencies,
due to 3-mode resonances. The results presented here possibly allows
for such an interpretation, as the amplitude of the ${\rm {}^2p_3-H_2}$ and
${\rm {}^2p_4-H_3}$ frequencies in Fig.~\ref{strain-high-band}, is larger
than the amplitude of the ${\rm {}^2p_2-H_1}$ frequency. Since the latter
involves modes of lower order (and of higher average kinetic energy in
our examples), one would expect the opposite result. It is thus
possible that the amplitude at the ${\rm {}^2p_3-H_2}$ and ${\rm {}^2p_4-H_3}$
frequencies is larger because of a resonance effect with the linear
${\rm {}^2f}$ mode. Although we have not included back-reaction effects, the
imprint of a possible resonance may be already present in the
structure of the equations at the bilinear order that is treated here.

In the present work, we have only focused on one particular, cold
equilibrium model. In a follow-up study, we will survey a larger
number of equilibrium modes and examine the dependence of the
detectability of bilinear harmonics in gravitational waves on the
equation of state and the compactness of the star. In a more realistic
study, the influence of the high entropy immediately after core
bounce, the subsequent thermal evolution of the equilibrium model and
rotational and magnetic-field effects should also be taken into
account. In addition, numerical simulations have shown that
significant mass accretion onto the proto-neutron star (PNS), the
settling of the PNS to smaller radii (factor of 2) and a non-neglible
density in the PNS hot envelope, all contributes to the gravitational
wave signal between the end of the initial phase of bounce ring down
($\sim 10$ ms post bounce) and the start of convection in the hot bubble
($\sim 70$ ms post bounce) \cite{Mueller2004}. When additional factors are
taken into account, our perturbative study could help in resolving
specific features in the complex gravitational wave signal in
realistic core-collapse.

In another study, we are planning on using amplitude
equations in perturbation theory and a fully nonlinear numerical code,
as in~\cite{Dimmelmeier:2005zk}, to also explore possible saturations
of the nonlinear oscillations or the existence of possible resonances
or parametric instabilities, as suggested
in~\cite{Passamonti:2004je,Passamonti:2005cz,Dimmelmeier:2005zk}.

%

\[ \] {\bf Acknowledgments:} 
We thank Marco Bruni, Massimo Cerdonio, Harry Dimmelmeier, Leonardo
Gualtieri, Michele Punturo, Carlos Sopuerta, Kostas Kokkotas and Ewald
M\"uller for fruitful discussions and comments on the manuscript.  We
are grateful to Massimo Cerdonio and Michele Bonaldi for providing the
sensitivity curve for the Dual SIC bar detector and to Michele Punturo
for the sensitivity curve of Advanced VIRGO.  A.P. is supported by a
``Virgo EGO Scientific Forum (VESF)'' grant and by the EU program
ILIAS.

\appendix

\section{Nonradial Perturbation Equations \label{sec:non-radial-eqs}}

In this section, we write the perturbation equations for some of the
linear nonradial perturbations of a nonrotating spherical star,
i.e. the metric variable~$\psi^{(0,1)}$, and the
velocity~$\gamma^{(0,1)}$ and $\alpha^{(0,1)}$.  In fact, the closed
system of perturbation equations for the enthalpy~$H^{(0,1)}$ and the
two metric quantities~$S^{(0,1)}$ and~$k^{(0,1)}$ is already given
in~\cite{Passamonti:2004je, passamonti-2006}.
\begin{align}
& \psi_{, \, t} ^{(0,1)}     =  - 2 \Phi_{, \, r} e^{\Phi - \Lambda} \left( r S^{(0,1)} + k^{(0,1)} \right) 
                            -  e^{\Phi + \Lambda} \left( r S^{(0,1)} \right)_{, \, r} \, , \\ 
& \gamma_{, \, t} ^{(0,1)}   =  - \Phi_{, \, r} \left( r S^{(0,1)} + k^{(0,1)} \right) 
                            - H_{, \, r} ^{(0,1)} + \frac{1}{2}  k_{, \, r} ^{(0,1)}  \, , \\
& \alpha_{, \, t} ^{(0,1)}   =   \left[ \frac{1}{2} \left(r S^{(0,1)} + k^{(0,1)} \right) - H^{(0,1)}  \right] e^{\Phi } \, , 
\end{align}
where from one of the TOV equation we have:
\begin{equation}
\Phi_{, \, r}  = \left( 4 \pi p r + \frac{m}{r^2} \right) e^{2 \Lambda} \, . 
\end{equation}

\section{Pulsational Kinetic Energy \label{sec:Kin_Energy}}
The kinetic pulsation energy can be determined as in
\cite{Thorne:1969to}, that in terms of the GSGM variables is given by
\begin{eqnarray}
E^{(0,1)} & = & \frac{1}{2} \int_{0}^{R} dr \left( \varepsilon + p \right) 
                \delta u_i \delta u^{i}  
                e^{\Phi} \sqrt{{}^3g} d^3 x  \, ,  \label{def_Kin_ener_01} 
\end{eqnarray}
where $d^3 x = dr \, d \theta \, d \phi$ and ${}^3 \, g$ is the
determinant of the 3-metric. The covariant 3-velocity of the
nonradial perturbations is instead given by the following expression:
\begin{equation}
\delta u_i = \left( \left( \tilde \gamma ^{(0,1)} -
              \frac{\psi^{(0,1)}}{2} \right) 
              e ^{\Lambda} Y^{lm} , 
              \tilde \alpha  ^{(0,1)} \, Y^{lm}_{, \, \theta} , 0 \right) \, , \label{3-vel}
\end{equation}
where $\tilde \gamma ^{(0,1)}$ and $\tilde \alpha ^{(0,1)}$ are two
functions that describes the radial and latitudinal velocity
components.  We have denoted them with a tilde because they are not
gauge invariant. However, in the Regge-Wheeler gauge we have that
$\tilde \gamma ^{(0,1)} = \gamma ^{(0,1)}$ and $\tilde \alpha ^{(0,1)}
= \alpha ^{(0,1)}$, where $\gamma ^{(0,1)}$ and $\alpha ^{(0,1)}$ are
the two polar gauge invariant velocity components~\cite{Gundlach:1999bt}.  
When we introduce Eq.~(\ref{3-vel}) into (\ref{def_Kin_ener_01}) and
calculate the integral over the 2-sphere, we obtain 
\begin{eqnarray}
E^{(0,1)} & = & \frac{1}{2} \int_{0}^{R} dr \left( \varepsilon + p \right) e^{\Phi+\Lambda} 
      \left[ r^2  \left( \tilde \gamma^{(0,1)} - \frac{\psi^{(0,1)}}{2} \right) ^2  \right.
      \nn \\   
          & + & \left. \ell  \left(\ell+1 \right) 
         \left( \tilde \alpha^{(0,1)} \right) ^2 \right] \label{Kin_ener_01_B} \, ,
\end{eqnarray}
that reduces to Eq.~(\ref{Kin_ener_01}) in the Regge-Wheeler gauge.

\nocite*

\begin{thebibliography}{60}
\bibitem[{\citenamefont{Adv-VIRGO}()}]{Ad_VIRGO_site}
\bibinfo{author}{\bibnamefont{Adv-VIRGO}},
  \bibinfo{note}{{wwwcascina.virgo.infn.it/advirgo}}.
\end{thebibliography}

\end{document}